# Pressure-induced phase transformation in zircon-type orthovanadate SmVO$_4$ from experiment and theory

C Popescu[1,*], Alka B Garg[2,†], D Errandonea[3], J A Sans[4], P Rodriguez-Hernández[5], S Radescu[5], A Muñoz[5], S N Achary[6] and A K Tyagi[6]

[1]*CELLS-ALBA Synchrotron Light facility, 08290 Cerdanyola, Barcelona, Spain.*

[2]*High Pressure and Synchrotron Radiation Physics Division, Bhabha Atomic Research Centre, Mumbai 400085, India.*

[3]*Departamento de Física Aplicada-ICMUV, MALTA Consolider Team, Universidad de Valencia, Edificio de Investigación, C/Dr. Moliner 50, Burjassot, 46100 Valencia, Spain*

[4]*Instituto de Diseño para la Fabricación y Producción Automatizada, MALTA Consolider Team, Universitat Politècnica de Valencia, 46022 Valencia, Spain*

[5]*Departamento de Física, Instituto de Materiales y Nanotecnología, MALTA Consolider Team, Universidad de La Laguna, La Laguna, E-38205 Tenerife, Spain*

[6]*Chemistry Division, Bhabha Atomic Research Centre, Trombay, Mumbai 400085, India*

* Corresponding author ; E-mail cpopescu@cells.es

† On deputation from BARC at ICMUV, Universidad de Valencia

**Abstract**

The compression behavior of zircon-type samarium orthovanadate, SmVO$_4$, has been investigated using synchrotron-based powder x-ray diffraction and *ab-initio* calculations up to 21 GPa. The results indicate the instability of ambient zircon phase at around 6 GPa, which transforms to a high-density scheelite-type phase. The high-pressure phase remains stable up to 21 GPa, the highest pressure reached in the present investigations. On pressure release, the scheelite phase is recovered. Crystal structure of high-pressure phase and equations of state (EOS) for the zircon- and scheelite-type phases have been determined. Various compressibilities such as bulk, axial and bond, estimated from the experimental data are found to be in good agreement with the results obtained from theoretical calculations. Calculated elastic constants show that the zircon structure becomes mechanically unstable beyond the transition pressure. Overall there is good agreement between experimental and theoretical findings.

Keywords: zircon, scheelite, x-ray diffraction, high pressure, phase transition, equation of state, vanadate

PACS numbers: 62.50.-p, 64.70.K-, 81.30.Hd, 64.10.+h

# 1. Introduction

Orthovanadates are materials of fundamental and technological importance due to a large variety of functional properties exhibited by them. These materials have potential applications such as scintillators, thermophosphors, photocatalyst, and cathodoluminescene[1]. In particular, as photocatalysts they have attracted great attention for their applications in renewable energy and alternative green technology[2]. They are also used as laser-host materials when doped with trivalent rare earth cation due to their high-optical conversion efficiency, high birefringence and good thermal conductivity[3,4]. At low temperature, these materials show interesting structural and magnetic phase transformations[5] along with a few exhibiting Jahn-Teller distortion[6,7].

Similar to the majority of rare-earth orthovanadates ($RVO_4$; R is rare-earth element), samarium vanadate, $SmVO_4$, crystallizes in tetragonal zircon-type structure (space group: $I4_1/amd$, $Z = 4$)[8,9]. In this structure, the vanadium atom is tetrahedrally coordinated while the samarium cation is coordinated by eight oxygen atoms, forming triangular dodecahedra (bidisphenoid). The structure can be described in terms of alternating edge-sharing $SmO_8$ dodecahedra and $VO_4$ tetrahedra forming chains parallel to the $c$-axis as shown in Fig. 1. Upon compression these compounds undergo transformations to different denser phases depending on the size of the rare-earth cation[10]. Typically, the zircon-type vanadates with small rare-earth cations transform to the tetragonal scheelite-type structure (space group: $I4_1/a$, $Z = 4$) around 6-8 GPa[11,12]. It has been shown that the zircon and scheelite structures are closely and simply related via crystallographic twin operation[13]. In particular, starting with zircon and twinning on (200), (020) and (002) generates the scheelite structure. Because of these symmetry relations the axial ratio of zircon ($c / a \approx 0.9$) is approximately equal to $2a / c$ in scheelite (i.e $c / a \approx 2.2$)[10]. On the other hand, in the compounds with larger rare-earth cations (La-Pr), a different structural sequence has been reported. These vanadates transform, under quasi-hydrostatic conditions to the monoclinic monazite-type structure (space group: $P2_1/n$, $Z = 4$) at similar pressures[14]. However, when compressed non-hydrostatically, the zircon to scheelite transition is observed[15]. Both non-hydrostatic stresses and the cationic radii have been shown to play similar role in the structural behavior of zircon-type arsenates, chromates and phosphates[16,17]. Raman spectroscopic measurements reported earlier in $SmVO_4$ indicates zircon to scheelite transition at 6.5 GPa[18]. However, to the best of our knowledge, $SmVO_4$ has not yet been studied under compression either by *in-situ* x-ray diffraction (XRD) or *ab-initio* calculations. These two techniques combined, have shown to be a powerful tool to understand the high-pressure behavior of ternary oxides[19]. In addition to the high-pressure structural behavior of SmVO$_4$, another issue that deserves to be studied is the possible decomposition of rare-earth orthovanadates under high-pressure when studied with long-wavelength x-rays, a phenomenon that has been reported to occur in HoVO$_4$ and EuVO$_4$[12,20].

In the present article, we report synchrotron based powder x-ray diffraction study of SmVO$_4$ using two different x-ray energies and *ab-initio* calculations up to the pressure of about 21 GPa. Evidence of zircon to scheelite phase transition is presented. The structural details of the low- and high-pressure phases have been determined at various pressures. Furthermore, the pressure dependence of unit-cell parameters, interatomic bond distances, elastic constants, and equations of state (EOS) of different phases are obtained. The reported results are discussed in comparison with related orthovanadates.

## 2. Experimental

Polycrystalline SmVO$_4$ was synthesized by a conventional solid-state reaction of stoichiometric amounts of pre-dried Sm$_2$O$_3$ and V$_2$O$_5$. Details on sample preparation can be found elsewhere[21]. The sample was characterized by powder XRD using a Panalytical X-pert Pro diffractometer employing Cu K$_\alpha$ ($\lambda$ = 1.5418 Å) radiation. A single phase with zircon-type structure was confirmed with unit-cell parameters as *a* = 7.2618(2) Å, *c* = 6.3837(3) Å and *V* = 336.64(3) Å$^3$ which agrees well with those reported in the literature[1,9].

Two series of experiments were performed at room temperature (RT) upon compression: run 1 up to 21.4 GPa and run 2 up to 20 GPa. The sample was loaded in 150 μm diameter hole of stainless steel or inconel gaskets inside of a Mao-Bell /membrane type diamond-anvil cell (DAC) with a diamond culet of 400 μm. A mixture of methanol-ethanol (ME) in 4:1 ratio was employed as pressure-transmitting medium (PTM) in run 1 whereas 16:3:1 methanol-ethanol-water (MEW) mixture was used in run 2. High-pressure angle-dispersive powder XRD (ADXRD) experiments were carried out at the XRD1 beamline of Elettra synchrotron source for run 1 and at the MSPD beamline of the Spanish ALBA synchrotron source[22] for run 2. At Elettra, monochromatic x-rays of wavelength 0.653 Å were used with a beam size limited to 80 μm in diameter by a circular collimator. The EOS of Pt was used as *in-situ* pressure scale[23]. A MAR345 image-plate area detector was used to collect XRD patterns for an exposure time of 15-20 minutes at each pressure. At ALBA, a monochromatic beam of wavelength 0.4246 Å was focused to a 15 μm×15 μm spot (full-width half maximum) using Kirkpatrick-Baez mirrors. Pressure was determined using Cu as internal pressure gauge[23].

Diffraction images were collected using Rayonix CCD detector with an exposure time of 10-30 seconds. The FIT2D software[24] was used to calibrate sample to detector distance, detector tilt and to integrate the two-dimensional diffraction images to standard one dimensional intensity *vs.* two-theta plot. The structural analysis was performed with GSAS software package[25] using a pseudo-Voigt profile function of Thompson, Cox, and Hastings[26]. The background of the XRD patterns was modeled with a Chebyshev polynomial function.

## 3. Theoretical calculations

*Ab-initio* total-energy calculations were performed within the framework of density-functional theory (DFT)[27] using the plan-wave method and pseudopotential theory with the Vienna *ab-initio* simulation package (VASP)[28,29,30]. The projector-augmented wave scheme[31,32] was employed in order to include the full nodal character of the all electron charge density in the core region. Basis sets including plane waves up to an energy cutoff of 520 eV were used to achieve highly converged results and an accurate description of the electronic properties. The exchange-correlation energy was considered in the generalized-gradient approximation (GGA) with the Perdew-Burke-Ernzerhof exchange-correlation prescription (PBE)[33]. We also performed calculations with PBEsol[34], however the results are very similar to the ones obtained with PBE. A dense Monkhorst-Pack grid of k-special points was used to perform integrations along the Brillouin zone (BZ) in order to obtain very well converged energies and forces. At each selected volume, the structures were fully relaxed to their equilibrium configuration through the calculation of the forces on atoms and the stress tensor. In the relaxed configurations, the forces on the atoms were smaller than 0.006 eV/Å and deviations of the stress tensor from a diagonal hydrostatic form are less than 0.1 GPa. Therefore our *ab-initio* calculations provide a set of accurate energy, volume and pressure (E, V, P) data that can be fitted using an equation of state (EOS) in order to obtain the equilibrium volume ($V_0$), bulk modulus (B) and its pressure derivate (B'). Calculations also provide the enthalpy of the different structures versus pressure, which allows us to determine the stable structure and the transition pressure.

Elastic constants describe the mechanical properties and mechanical stability of materials in the region where the stress-strain relations are still linear. The elastic constants were obtained by computing the macroscopic stress with the use of the stress theorem[35]. The ground state and fully relaxed structures at several pressures were strained according to their symmetry. The small total-energy differences between the different strained states were evaluated with high precision according to a Taylor expansion[36] of the total energy with respect to the applied strain. Therefore it is important to check that the strain used in the calculations guarantees the harmonic behavior.

Compound crystallizing in the zircon type structure belongs to the tetragonal Laue group TI which has six independent elastic constants $C_{11}$, $C_{12}$, $C_{13}$, $C_{33}$, $C_{44}$ and $C_{66}$. On the other hand, the scheelite-type structure belongs to the tetragonal Laue group TII and has seven elastic constants $C_{11}$, $C_{12}$, $C_{13}$, $C_{33}$, $C_{44}$, $C_{66}$ and $C_{16}$. To analyze the mechanical stability of crystals under external hydrostatic pressure P, the "Generalized Born Stability Criteria" must be applied[37]. For the TI Laue group these criteria take the form:

$$M1 = C_{11} - P > 0$$

$$M2 = C_{11} - P - |C_{12} + P| > 0$$

$$M3 = (C_{33} - P)(C_{11} + C_{12}) - 2(C_{13} + P)^2$$

$$M4 = C_{44} - P > 0$$

$$M5 = C_{66} - P > 0$$

For a crystal to be mechanically stable, all the above criteria should be satisfied simultaneously.
For a crystal belonging to the scheelite-type structure (TII Laue group), the stability criteria M1 to M4 are the same as for TI, however the M5 criterion must be replaced by the M6 criterion:

$$M6 = (C_{66} - P)(C_{11} - C_{12} - 2P) - 2(C_{16})^2 > 0$$

Note that when no hydrostatic pressure is applied (P = 0 GPa) the stability criteria M1 to M6 are known as "Born Stability Criteria"[38].

Lattice-dynamic calculations were performed at the zone center (Γ point) of the BZ with the direct force constant approach[39]. The construction of the dynamical matrix implies separate calculations of the forces in which fixed displacements from the equilibrium configuration of the atoms within the primitive cell is considered. Highly converged results for forces are required for the calculation of the dynamical matrix. Diagonalization of the dynamical matrix provides the frequencies of the normal modes.

# 4. Results and discussion

A selection of ADXRD patterns collected at several pressures using ME as PTM are shown in Figs. 2(a) & (b). A few data points measured while unloading pressure are also shown in Fig. 2 (c). In this set of data the ambient zircon phase is found to remain stable till 4.2 GPa. The data collected at 4.8 GPa shows the appearance of new peaks indicting the onset of structural phase transition in the material. On further compression, a few more diffraction peaks appear as the phase fraction of the HP phase increases while the peaks from the zircon phase reduces in intensity as its fraction goes down. The position of the strongest peak of the new structure matches well with that of the expected strongest peak of scheelite phase and indeed we could refine the new structure with a scheelite-type structure. At 10 GPa a pure scheelite phase could be observed which then remains stable up to 21.4 GPa, the highest pressure reached in this experiment. On decompression the scheelite phase is found to be metastable, being possible to recover it at ambient pressure. Another interesting aspect observed in this study was the appearance of a peak at 2θ value of ~12° at 6 GPa. This observation is similar to the one observed by us in other orthovanadates like $HoVO_4$ and $EuVO_4$[12,20]. The position of the peak matched well with that of the strongest peak of $V_2O_5$ and it could be followed up to the complete release of the pressure. The presence of $V_2O_5$ can be explained by a partial decomposition of the sample under the combined effect of compression and x-ray energy. In our earlier HP studies on $HoVO_4$ and $EuVO_4$[12,20], the appearance of the $V_2O_5$ peaks have been observed when the x-ray of wavelength close to 0.6 Å was employed. In the measurements made on these compounds with smaller or larger wavelengths the decomposition was not observed. Also decomposition was not observed in Raman experiments[18]. These facts support the idea that partial decomposition could be triggered by x-ray absorption which could induce photoelectric processes leading to the dissociation of $V_2O_5$ units from the vanadates[40].

In Fig. 3(a) we show the Rietveld refined pattern for the zircon structure as measured at ambient pressure in the DAC. Fig. 3(b) depicts the multiphase refinement consisting of zircon, scheelite and $V_2O_5$, whereas Figs. 3 (c) & (d) show the pure scheelite and pressure released scheelite phases, respectively. Since the phase fraction of the $V_2O_5$ was low only Lebail fitting was carried out as shown in Fig. 3(b) & (c), however for the zircon and scheelite phases Rietveld refinement was done. The low intensity of the $V_2O_5$ peaks facilitates the multiphase refinement of the two structures of $SmVO_4$. The structural parameters for the zircon phase at ambient pressure obtained from this data are $a = 7.2680(3)$ Å, $c = 6.3914(3)$ Å and $V = 337.79(4)$ Å$^3$ with various residuals of the refinement as $R_{WP} = 3.9$ % and $R_P = 2.82$%. For the scheelite phase at 10 GPa the structural parameters are $a = 5.044(1)$ Å, $c = 11.275(5)$ Å and V = 286.9(2) Å$^3$ with various residuals of the refinement as $R_{WP} = 2.4$ % and $R_P = 1.8$%.

To check the effect of x-ray wavelength on partial decomposition and also the effect of experimental conditions on the compressibility behavior of the zircon and scheelite phases, we have carried out another set of experiment using shorter wavelength (0.4246 Å) with different PTM (MEW). Fig. 4 shows a selection of XRD patterns for SmVO$_4$ measured in run 2. In the figure, two Bragg peaks associated with Cu can be easily identified since it has a different pressure evolution compared to the sample peaks. At nearly ambient pressure, inside the DAC, the XRD patterns can be unequivocally assigned to the tetragonal zircon-type structure. One can see it in the Rietveld refinement shown in Fig. 5(a) with the corresponding residuals of the structure determination. The R-factors of the refinement are $R_{wp}=3.8$% and $R_p = 2.75$%. The unit-cell parameters determined at 0.05 GPa are $a = 7.2603(2)$ Å, $c = 6.3890(3)$ Å and $V = 336.77(3)$ Å$^3$. These values are comparable with those previously reported in the literature[1,9] and with the parameters determined from run 1. XRD patterns can be indexed with the initial zircon-type phase up to 6.7 GPa. Above this pressure new Bragg peaks emerge which clearly indicates the existence of a pressure-induced phase transition in SmVO$_4$.

Upon further compression, the peaks of the low-pressure phase disappear completely at 9.7 GPa. Therefore the two phases, i.e. zircon and high-pressure phases, are found to co-exist from 6.7 to 8.7 GPa. The onset of the transition pressure is close to that previously reported in Raman measurements carried out using same PTM (onset at 6.5 GPa)[18]. At 9.7 GPa, the XRD peaks could be assigned to the scheelite structure as found in run 1. In this regard, SmVO$_4$ behaves in similar way as orthovanadates with smaller trivalent cations (LuVO$_4$, YbVO$_4$, EuVO$_4$, TbVO$_4$ and HoVO$_4$)[10,11,12,20,41]. We have assigned the scheelite-type structure to the HP phase as supported by structural refinements. The residuals of the refinement at 9.7 GPa are shown in Fig. 5(b). The R-factors are $R_{wp} = 3.3$% and $R_p = 2.3$%. The unit-cell parameters of the scheelite structure at this pressure are $a = 5.0019(7)$ Å, $c = 11.2341(9)$ Å and $V = 281.1(1)$ Å$^3$.

The scheelite structure of SmVO$_4$ remains stable up to ~ 20 GPa, the highest pressure reached in run 2. Upon decompression the phase transition is not reversible. When totally releasing the force applied to the DAC, a pressure of 0.7 GPa is achieved due to piston-cylinder friction in the DAC. At this pressure, the scheelite structure remains the stable phase as can be seen in Fig. 4. The unit-cell parameters of the scheelite-type $SmVO_4$ at 0.7 GPa are $a$ = 5.0992(6) Å, $c$ = 11.5115(11) Å and $V$ = 299.32(6) Å$^3$. The obtained unit-cell parameters are comparable with those reported in previous studies from quench experiments[18,42]. We would like to mention here that in run 1 the onset of transition occurs at a pressure ~ 2 GPa lower than in run 2 along with an increased pressure range of co-existence of both phases (~ 5 GPa). The most probable reason for this observation could be that the sample was bridged between the diamond anvils due to high occupancy of the sample in the gasket hole resulting in large deviatoric stresses in the sample well below its hydrostatic limit[43].

In order to study the effect of pressure on the crystal structure of $SmVO_4$ theoretically, in addition to zircon, we considered scheelite and monazite structures (most probable phases observed in $RVO_4$ compounds under compression). However, the only structures found to be stable are zircon and/or scheelite in the pressure range covered in our studies. Fig. 6 shows the total-energy *vs.* volume plot showing the stability of zircon structure at ambient pressure. These curves are used to determine the P-V (EOS) of the zircon and scheelite phases. Inset in Fig. 6 shows the enthalpy *vs.* pressure plots for zircon and scheelite structures. It is clear from the figure that zircon-type structure of $SmVO_4$ has the lowest enthalpy at ambient pressure. The calculated structural parameters at ambient pressure for the zircon structure are $a$ = 7.34974 Å, $c$ = 6.41562 Å and $V$ = 346.564 Å$^3$. These values agree within 1% with the experimental results. Upon compression calculations predict (from the enthalpy vs. pressure plot) the occurrence of a zircon to scheelite phase transition at 4.5 GPa. This transition pressure is in good agreement with the experimental values. It is a first-order transition that involves a large volume collapse ($\Delta V/V$ = 10 %). The calculated structural parameters at 7.9 GPa for the scheelite structure are $a$ = 5.06455 Å, $c$ = 11.35874 Å and $V$ = 291.34 Å$^3$. In table I the refined atomic positions for both the phases from two experiments along with the data obtained from theoretical calculations are given showing good agreement. However, the calculated lattice parameters slightly overestimate the experimental results which is usual with DFT calculations.

In Fig. 7, the unit-cell parameters for both the phases obtained from run 1, run 2, and theory are plotted along with the pressure released data. The compressibility of $a$ and $c$ axis of the zircon phase is comparable from run 1, run 2 and theory. However, there is large scatter in $c$ parameter of scheelite phase from ME data. The same effect was observed for $HoVO_4$[12], when experimental conditions induced large deviatoric stresses. This observation is compatible with the above mentioned comment on the possible sample bridging that might have occurred in run 1. As a consequence of it, the sample compressibility is usually reduced, as was found in the experimental data obtained with the same PTM for HoVO$_4$[12]. Another fact that deserves to be mentioned from Fig. 7 is that the *c*-axis is less compressible than the *a*-axis. As a consequence of it, the axial ratio *c/a* is increased by 2% from ambient pressure to the transition pressure. In the scheelite structure, the reverse behavior is observed. The most compressible axis is the *c*-axis. Then, the *c/a* ratio decreases under compression, about a 2% from the transition pressure to 20 GPa. The same anisotropic behavior has been observed for both phases in related compounds which is associated with the different polyhedral chains present in each structure and to the fact that the SmO$_8$ units are more compressible than the VO$_4$ polyhedra[15].

Fig. 8 depicts the experimental pressure volume data for both the phases from run 1, run 2 along with theoretical results. The obtained ambient pressure bulk modulus for zircon and scheelite phases by fitting the experimental data from both the runs to the 2$^{nd}$ order Birch-Murnaghan EOS along with the theoretical EOS are given in table II. The value of bulk modulus obtained from both runs agrees well with the theoretical results for zircon phase. The values of the bulk moduli are also comparable with the values reported for the zircon phase of other vanadates[10,11,12,14,15,20,41]. For the scheelite phase there is a good match between the values obtained from run 2 and theory but the value obtained from run 1 is extremely large. Since the hydrostatic limit of the PTM used in both runs is almost same, we think that it is the effect of large amount of sample in the pressure chamber as mentioned earlier, which in turn makes the compression of the sample almost like non-hydrostatic leading to the observed large bulk modulus[12,43]. In fact deviatoric stresses have been shown to considerably influence the HP structural behavior of the isomorphic HoVO$_4$, CeVO$_4$ and EuVO$_4$[1,12,20]. These studies clearly indicate the importance of sample to PTM volume filling ratio in the gasket hole to estimate the correct compressibility behavior of materials under investigation.

Finally, from the Rietveld refined data from run 2 and calculations we extract the information on bond compressibility. Similar evolution for the bond distances are found from the experiment and calculations as depicted in Fig. 9. In both the phases the most compressible bonds are Sm-O bonds. In addition, after the phase transition, the V-O bond distance is slightly enlarged, increasing the volume of VO$_4$ tetrahedra. In case of Sm-O bonds, the longer bond reduces at the transition whereas the shorter Sm-O bond increases, making the SmO$_8$ dodecahedra more regular. However, the overall effect of compression on SmO$_8$ polyhedral unit is to reduce its volume. Both facts are a consequence of atomic rearrangement associated with the first-order zircon to scheelite phase transition, which is reconstructive and involve the breaking of interatomic bonds and formation of new bonds. In both phases the changes of the atomic bond is driven not only by the reduction of the lattice parameters when the material is squeezed but also by the movement of the oxygen atoms. In zircon the internal position of the oxygen atom changes from (0, 0.43132, 0.20306) at ambient pressure to (0, 0.43310, 0.20269) at 5 GPa. In the scheelite phase the internal position of oxygen changes from (0.24772, 0.10938, 0.04460) at 5 GPa to (0.24942, 0.10169, 0.04171) at the highest pressure. However, the changes in the atomic coordinates of oxygen are only in the third digit. This suggest that our assumption, that atomic positions are not affected much by pressure, as done in the past[10], is not a bad approximation for estimating the bond distances for the zircon and scheelite phases of orthovanadates.

The elastic constants, $C_{ij}$, of $SmVO_4$ for the low pressure zircon-type structure at P = 0 GPa are given in table III. These values matches well with the average values of other zircon-type ortovanadates[44]. The pressure evolution of the elastic constants is plotted in Fig. 10. Although $C_{44}$ is almost constant, the other elastic constants increase linearly with pressure, with the exception of $C_{66}$ which exhibits an inverse relation with pressure. It must be noted that, though $C_{11} < C_{33}$ the pressure derivative of $C_{11}$ is almost twice as that of $C_{33}$. This behavior of elastic constants has been described earlier for zircon $(ZrSiO_4)$[45]. The softening of $C_{66}$ is related to the pressure induced phase transition to the scheelite-type structure. As one can see in the inset of Fig. 10, the M1 to M5 criterion are fulfilled below 7.5 GPa and at 7.5 GPa the M5 stability criterion for the zircon-type structure is violated which indicates the mechanical instability of zircon structure at 7.5 GPa. This pressure is an upper limit for the phase transition. The result is in agreement with the theoretical and experimental findings reported here. Furthermore, the pressure evolution of silent mode $B_{1u}$, plotted in Fig.11, shows the softening and becomes imaginary at the same pressure, indicating the dynamical instability of the zircon-type $SmVO_4$ at 7.8 GPa.

The seven elastic constants calculated for the scheelite-type structure of $SmVO_4$ at ambient pressure are presented in Table III. As shown in Fig. 12, except for $C_{66}$, all the elastic constants increase with pressure. The negative evolution of the pressure coefficient of $C_{66}$ is again related with the violation of stability criterion. In this case the M6 criterion which involves $C_{66}$ and $C_{16}$, not present for the zircon-type, is not fulfilled at 24 GPa (inset in Fig. 12). This is the upper limit of the pressure beyond which the scheelite phase will be mechanically unstable. Beside this, the acoustic infrared mode $A_u$ softens and becomes imaginary at 22 GPa thus the scheelite–type structure of $SmVO_4$ is dynamically unstable at this pressure. This results suggests that further experimental studies, beyond the pressure limit covered in this work, are needed to identify the post-scheelite structure of $SmVO_4$.

Before concluding, we would like to mention that the bulk modulus for both tetragonal structures can be obtained from the combination of the elastic constants[46]. In our case the values obtained, 120.9 GPa for the zircon-type and 140.0 GPa for the scheelite type are in good agreement with those obtained from the calculated EOS. This indicates the quality and consistency of our calculations.

## 5. Conclusions

The compression behavior of $SmVO_4$ is investigated using synchrotron based powder x-ray diffraction along with *ab-initio* calculations. A structural phase transition is observed in both the experiments which corroborates well with theoretical findings. The present study provides quantitative information on structural parameters and equation of state of the zircon and scheelite phases. The compression of both phases is found to be anisotropic. The elastic constants for both the phases have also been calculated at different pressures. These calculations support the existence of a zircon-scheelite transition, indicating that the zircon structure becomes mechanically unstable after the transition pressure. Finally, the possible decomposition of orthovanadates when long x-ray wavelengths are used is confirmed. The obtained results are discussed in comparison with the known behavior of related compounds. The reliability of the present experimental and theoretical results is supported by the consistency between the results yielded by both techniques.

**Acknowledgments**

This work was partially supported by the Spanish MINECO under Grants MAT2013-46649-C04-01/02 and MAT2015-71070-REDC. Alka B. Garg would like to acknowledge the Indian DST for travel support and Italian DST for hospitality at Elettra. She also acknowledges fruitful discussions with Dr. Surinder M. Sharma.

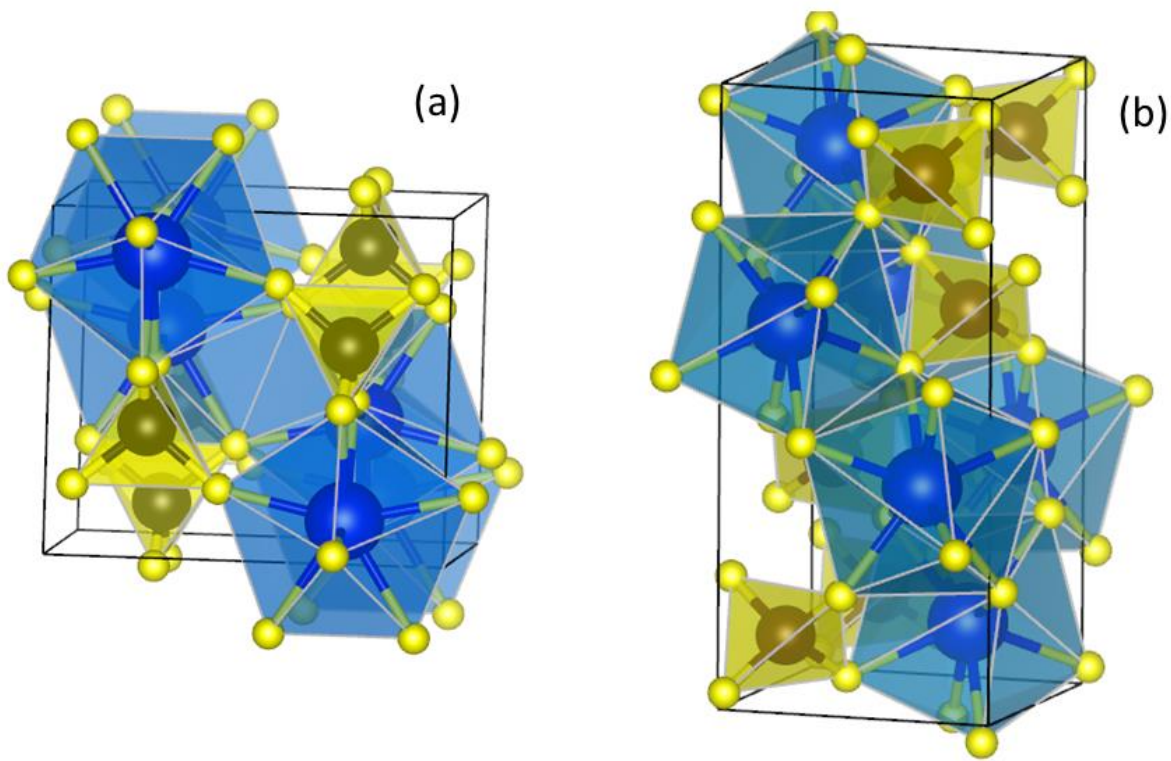

Figure 1: Crystal structure of zircon and scheelite SmVO$_4$. The large blue spheres are Sm atoms and medium-size black spheres are V atoms. The small yellow spheres are O atoms. (For interpretation of the references to color in this figure legend, the reader is referred to the web version of the article).

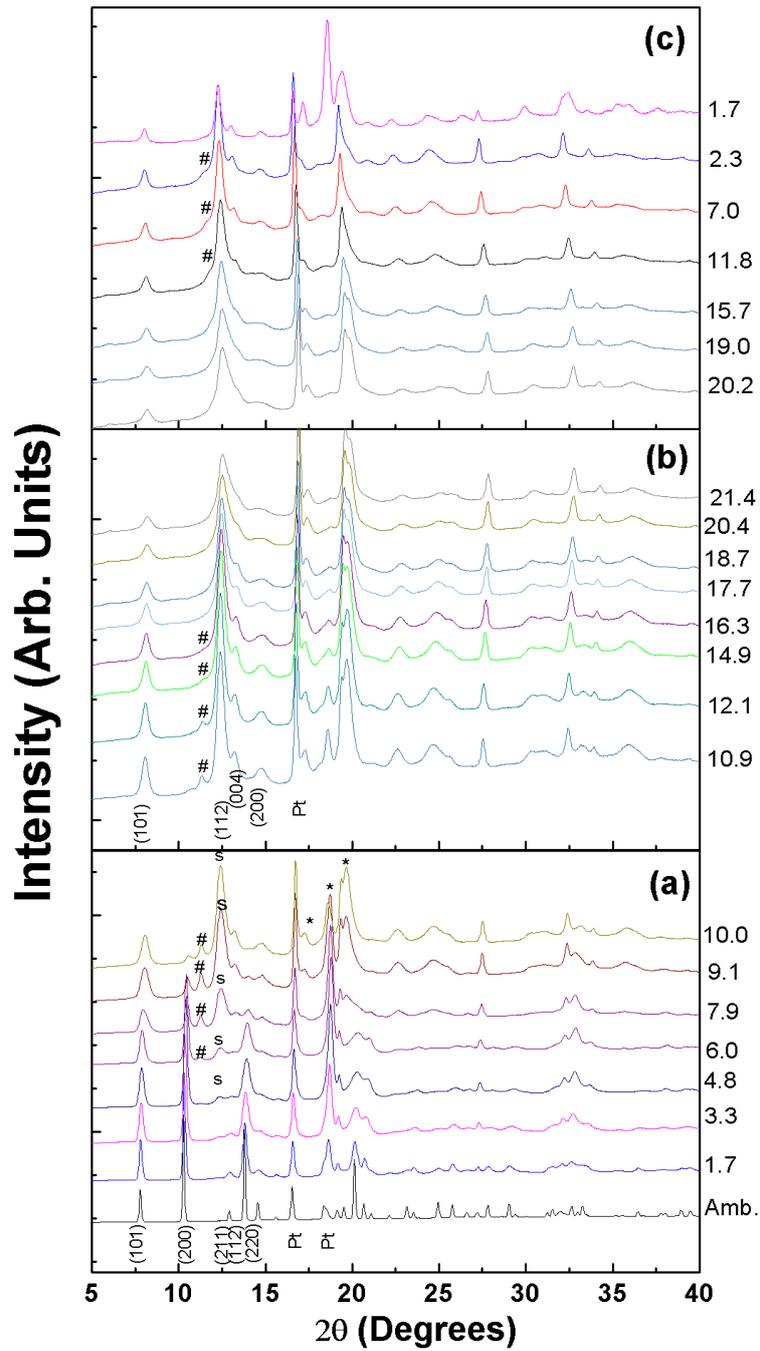

Figure 2: Evolution of XRD data with increasing pressure (a) & (b) from run 1 (PTM = ME). Panel (c) shows the diffraction patterns collected while unloading the pressure. Peaks from platinum pressure marker are marked by Pt. Pressures marked on the right hand side y-axis are in GPa. Gasket peaks are marked with *. Appearance of strongest scheelite peak is indicated by **s** whereas # denotes the strongest peak from the partially decomposed $V_2O_5$. For details see the text.

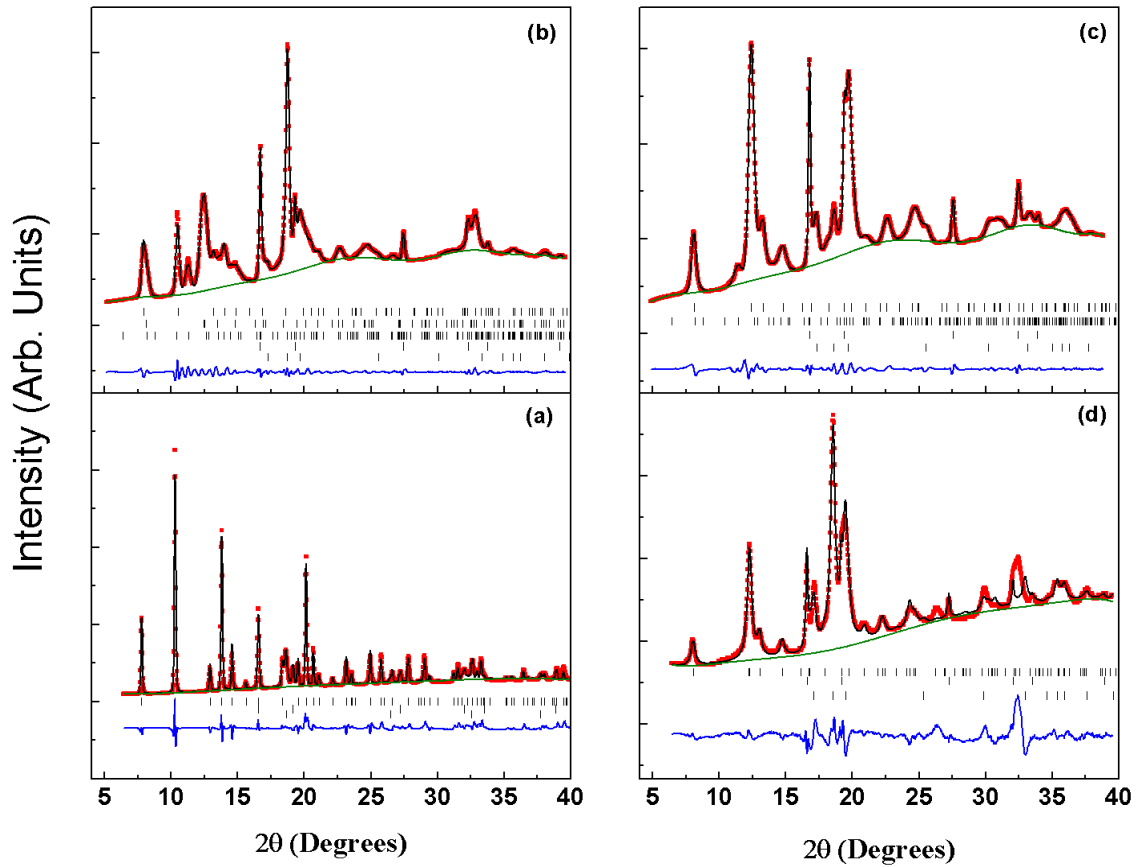

Figure 3: The Rietveld refinement of the various phases observed in run 1 (PTM = ME). Panel (a) shows the ambient zircon phase (top most vertical symbols are calculated peak positions from the zircon, middle are for the Pt pressure marker and the bottom tick marks are for the gasket). In panel (b) at 7.9 GPa, calculated peak positions from various phase such as zircon (top most vertical marks), scheelite (identified by the tick marks placed second from the top), $V_2O_5$ (ticks third from the top), pressure marker (ticks second from the bottom) and gasket (bottom tick mark) are shown. Panel (c) shows the pure scheelite phase while pressure loading (top most vertical ticks identify the scheelite peaks, second from the top are for $V_2O_5$, the pressure marker and gasket peaks are identified by the third from the top and the bottom symbols). Panel (d) shows the recovered scheelite phase (top most vertical ticks designate the scheelite peaks, middle are for the pressure marker and gasket peaks are identified by the bottom symbols). Residuals of the Rietveld refinement are also shown in each panel.

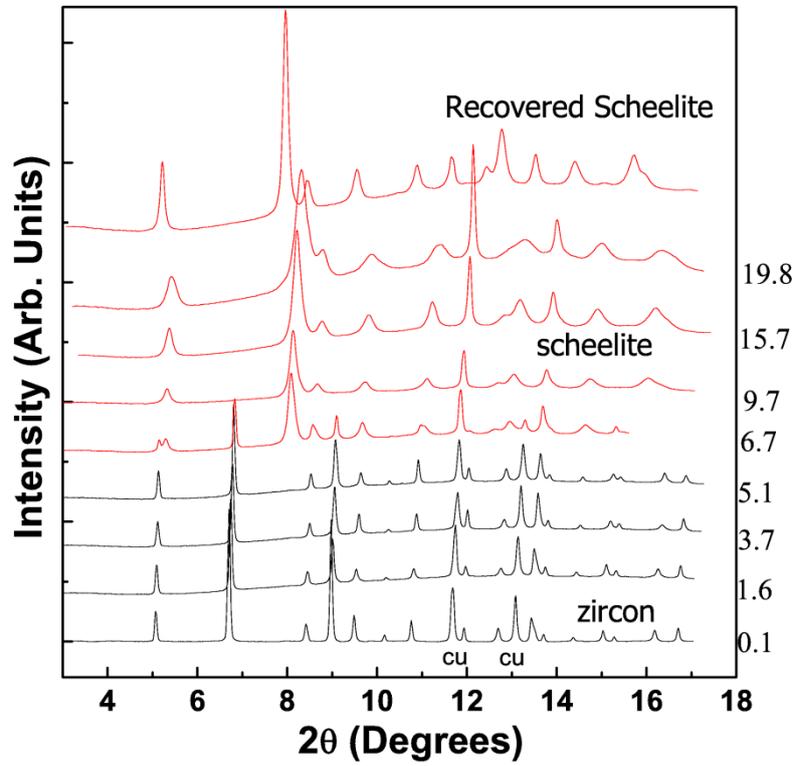

Figure 4: X-ray diffraction patterns at selected pressures measured in run 2 (PTM = MEW). The peaks of Cu (pressure gauge) are also identified. Pressures marked on the right hand side y-axis are in GPa. The pattern in the upper trace was collected at ambient pressure after decompression.

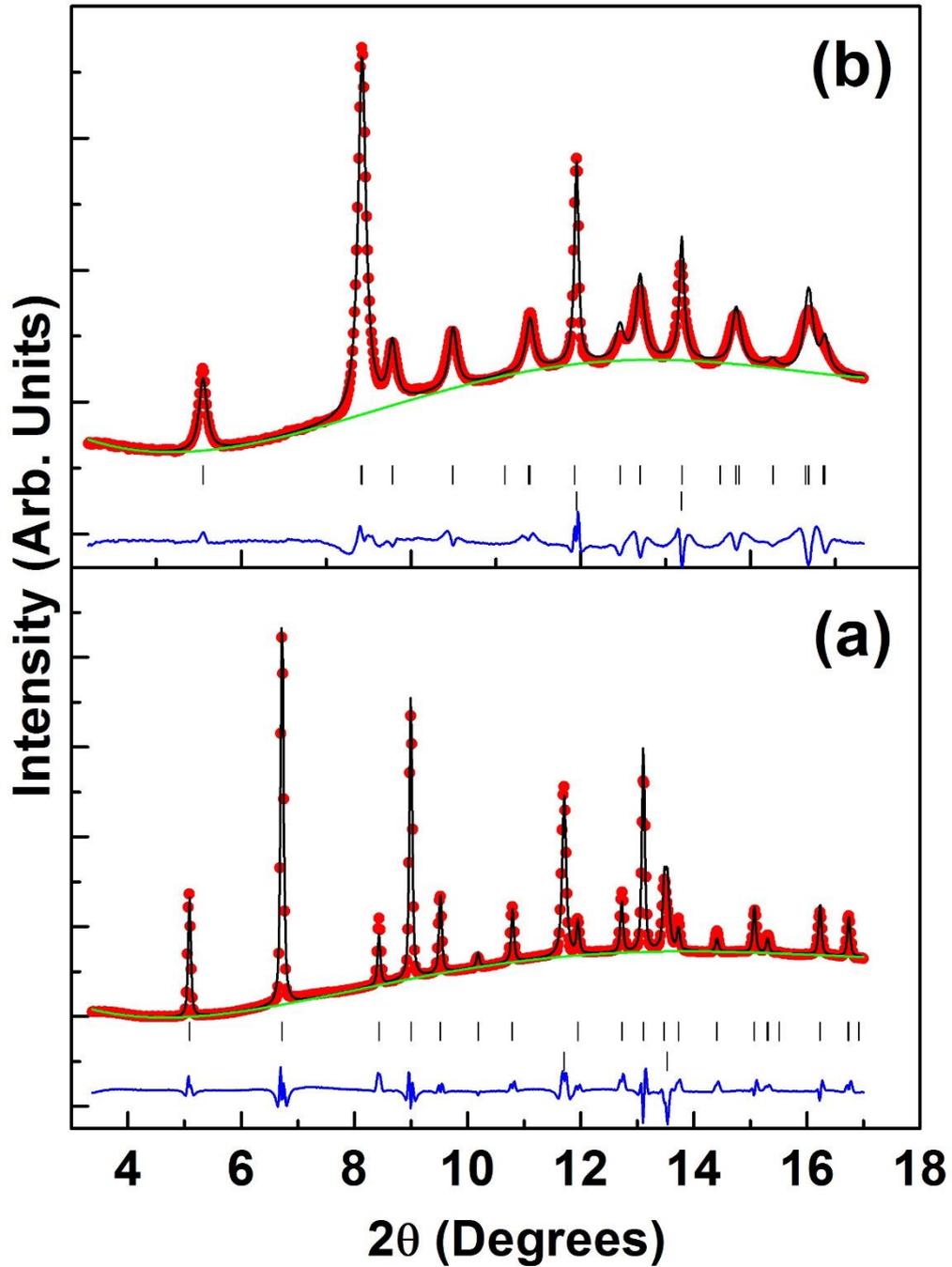

Figure 5: Rietveld refinements of selected patterns of SmVO$_4$. The residuals of the refinements are also shown. Panel (a) corresponds to the low-pressure phase (zircon) whereas panel (b) shows the high-pressure phase (scheelite). Upper ticks indicate the calculated positions of Bragg peaks for the two structures whereas the lower ticks identifies the peaks for the Cu (pressure marker).

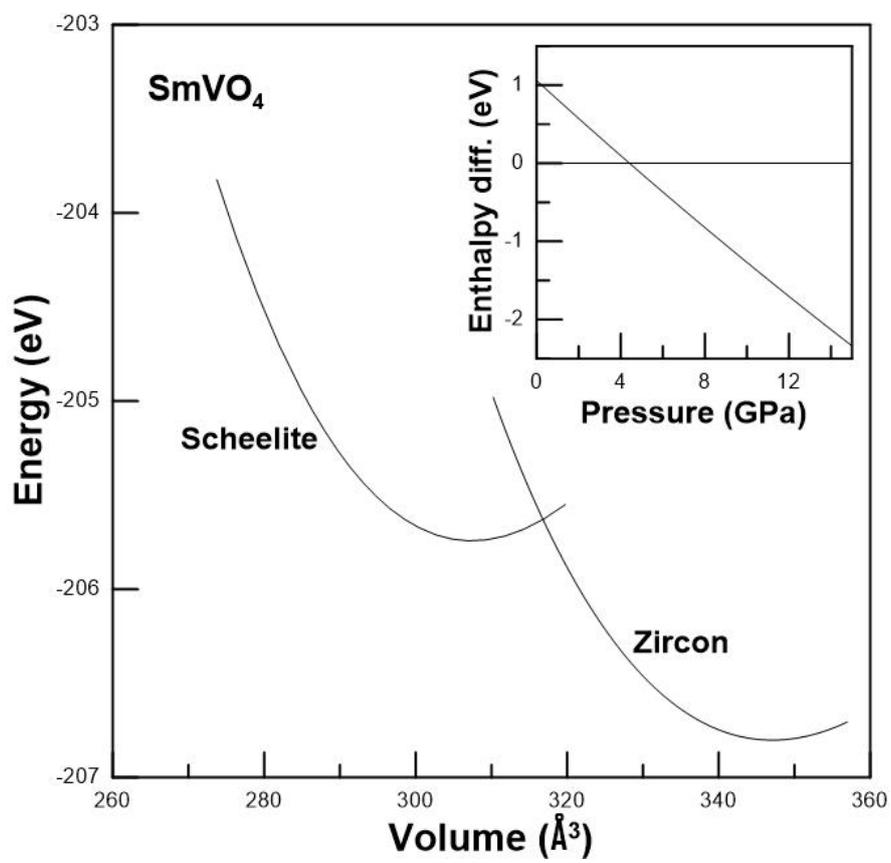

Figure 6: Total-energy versus volume from *ab-initio* calculations for the studied phases of SmVO$_4$. The inset shows the enthalpy as function of pressure, the zircon structure is taken as reference.

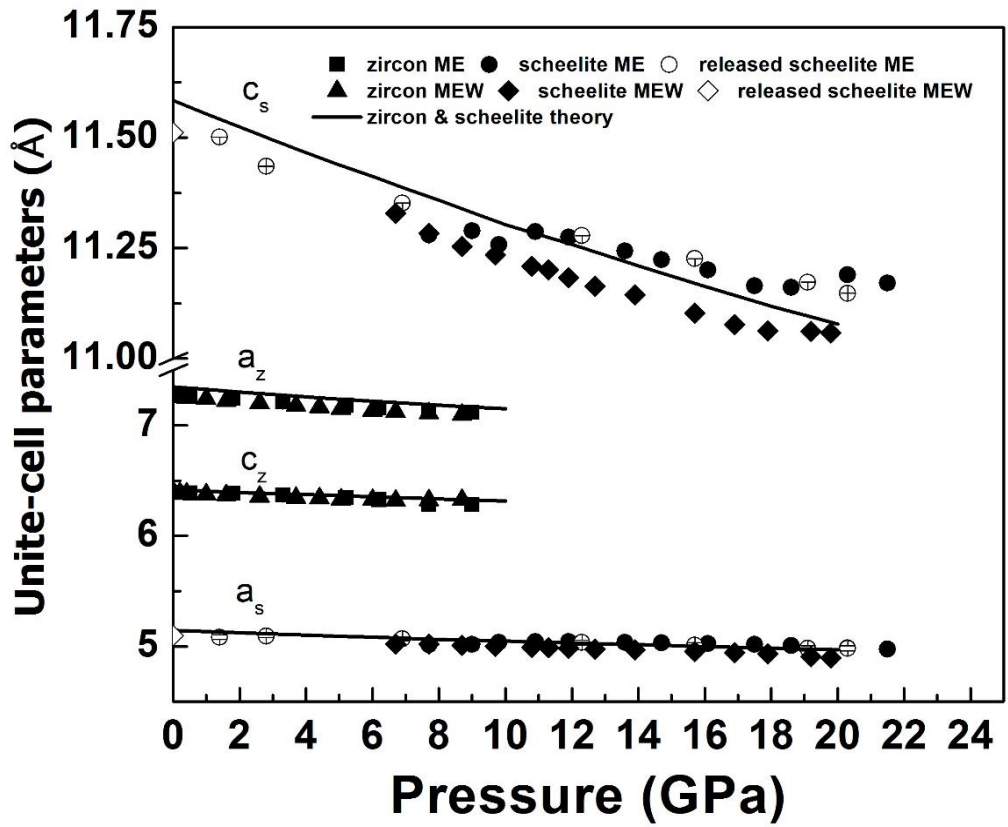

Figure 7: Unit-cell parameters for ambient zircon and high pressure scheelite phases obtained from two experimental runs and theory. Error bars have also been plotted for all the data sets. Symbols $a_z$ and $c_z$ refers to the *a* and *c* parameters for zircon phase while $a_s$ and $c_s$ refers to the *a* and *c* parameters for scheelite phase. Various phases have been identified in the figure.

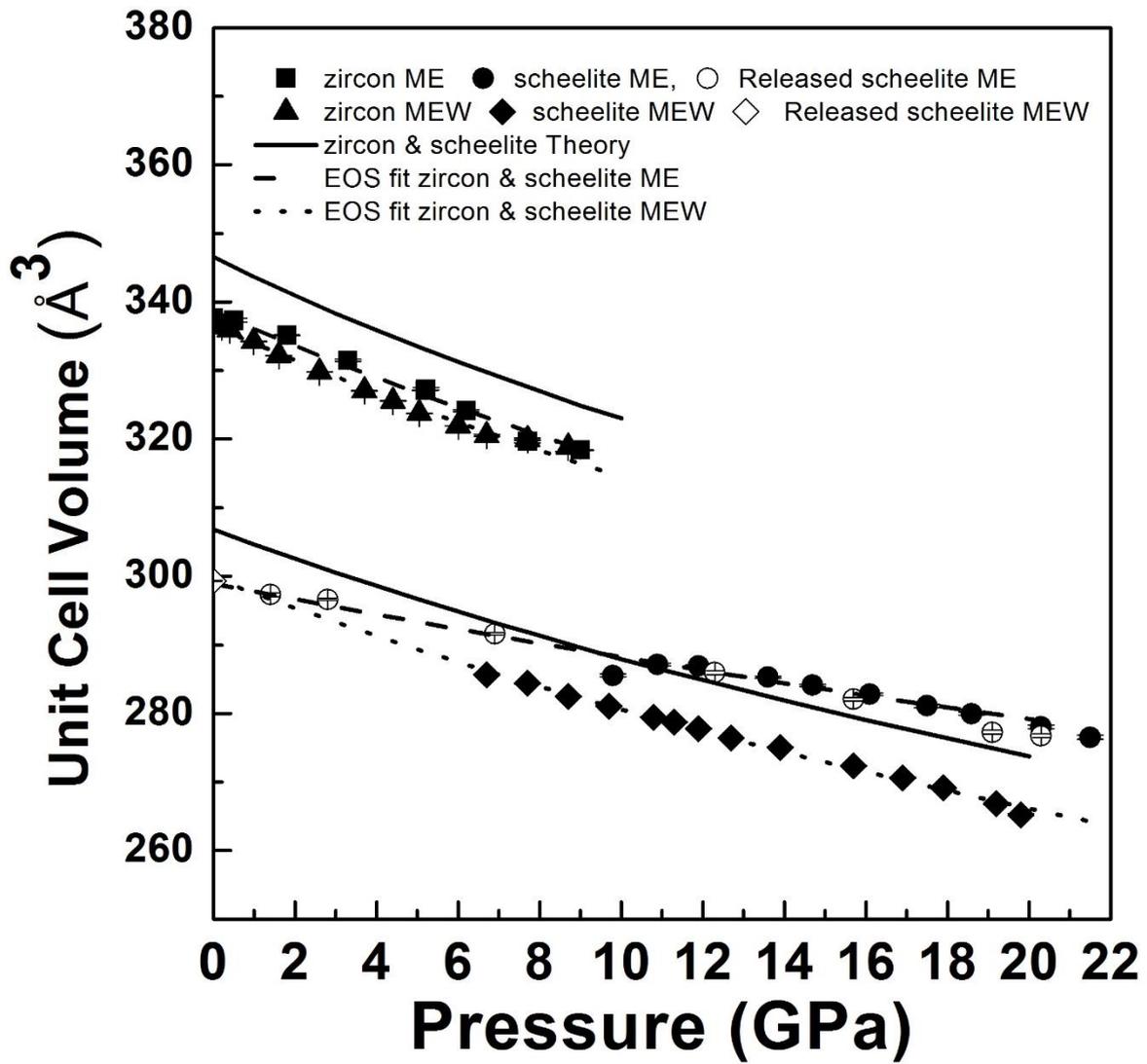

Figure 8: Unit-cell volume vs. pressure for both zircon and scheelite phases from two experimental runs along with theory. Various phases have been identified in the figure.

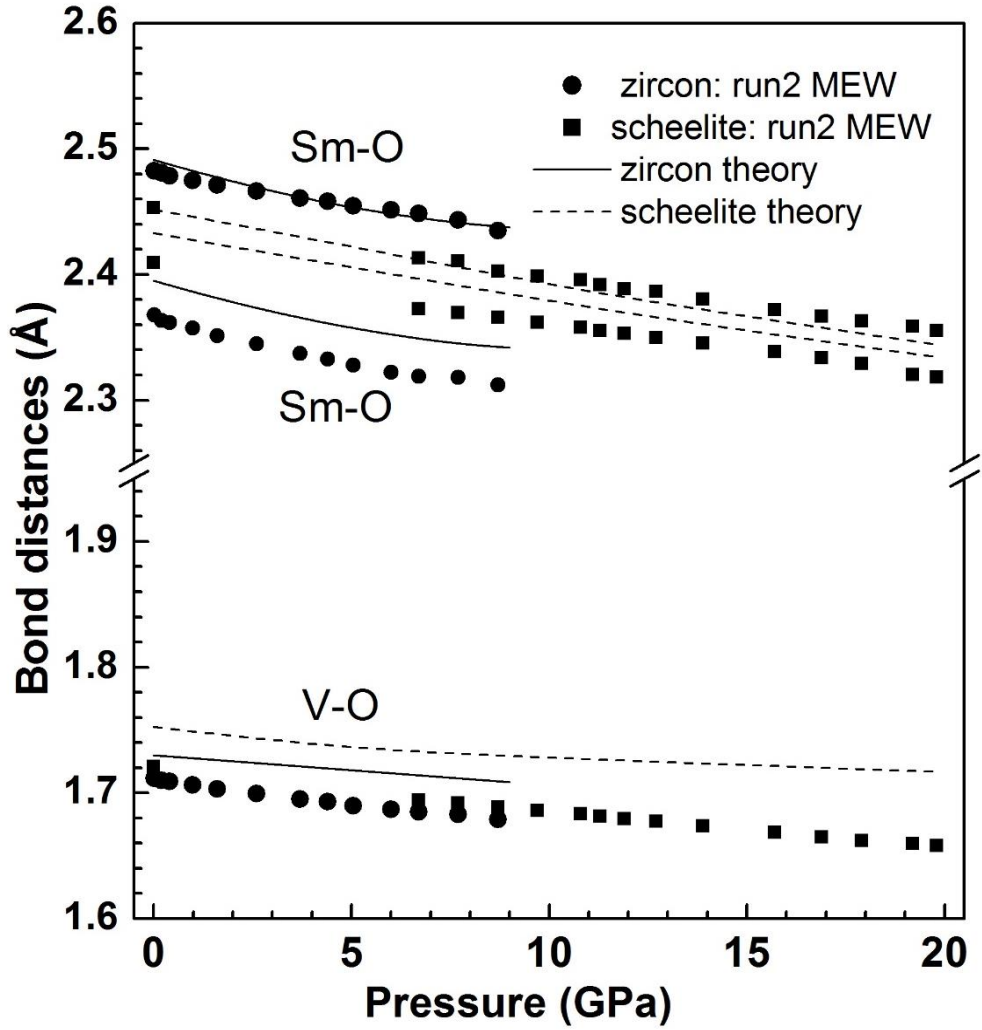

Figure 9: Various bond distances vs. pressure for both zircon and scheelite phases from experiment and theory.

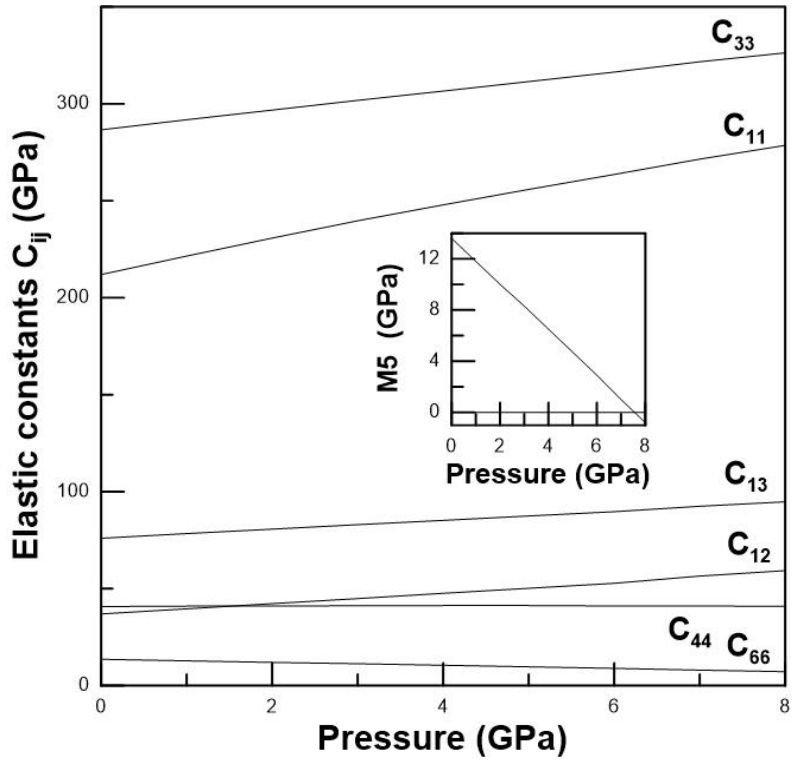

Figure 10: Pressure dependence of the six elastic constants of zircon-type structure of SmVO$_4$. Inset shows the M5 stability criterion versus pressure for the zircon-type structure.

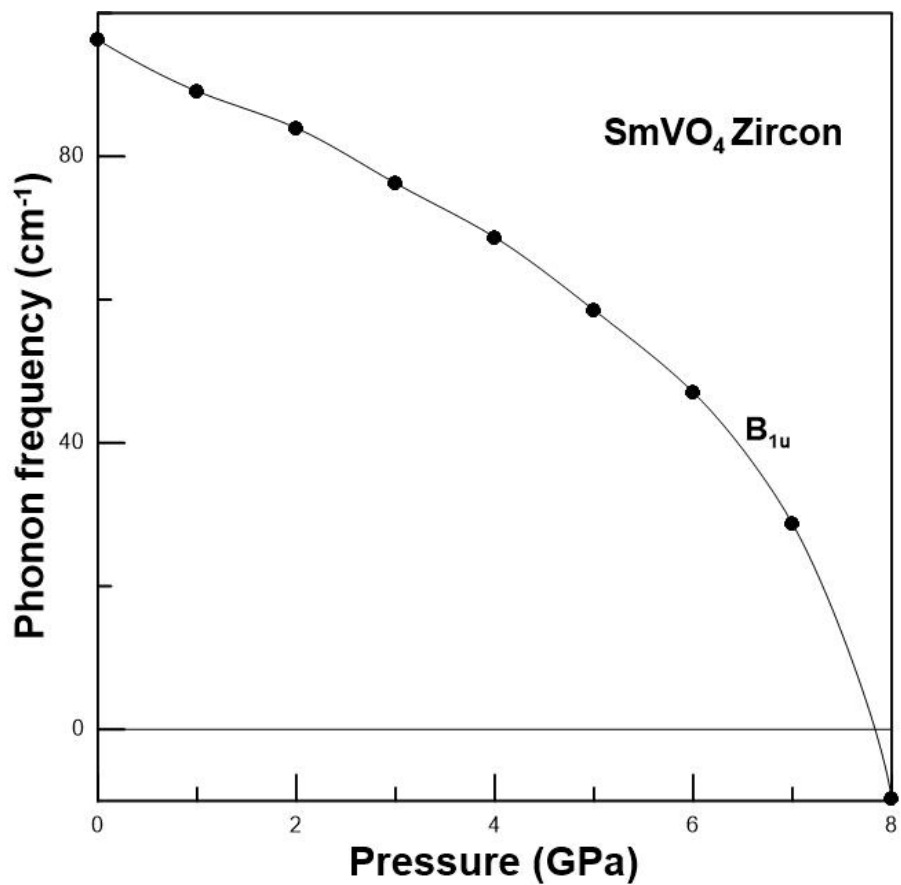

Figure 11: Pressure evolution of the silent phonon $B_{1u}$ of zircon-type structure of $SmVO_4$.

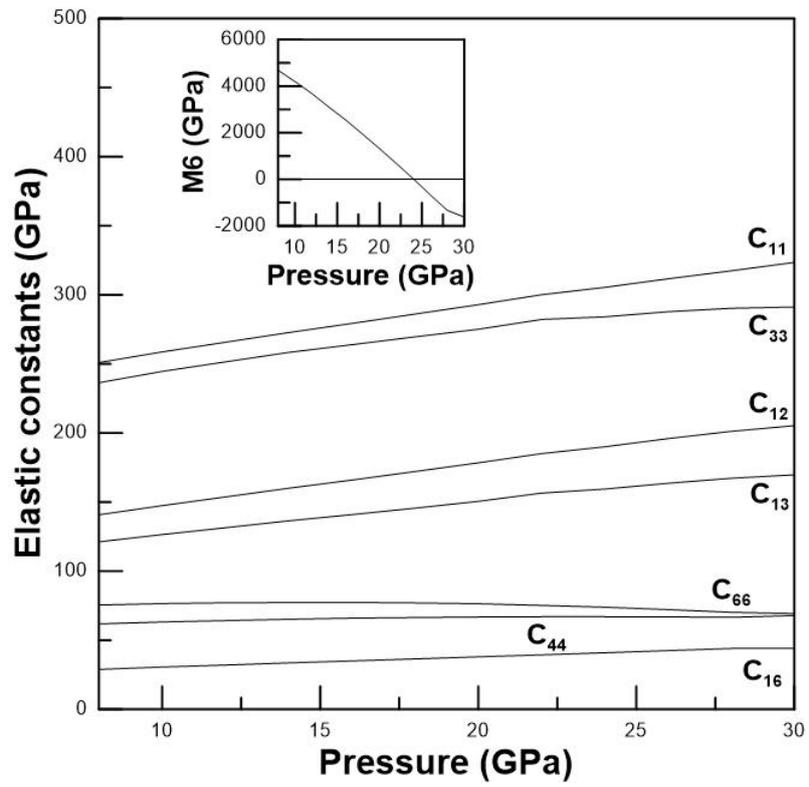

Figure 12: Pressure dependence of the seven elastic constants of scheelite-type structure of SmVO$_4$. Inset shows the M6 stability criterion versus pressure for the scheelite-type structure.

**Table I.** Refined atomic positions from XRD patterns measured in run 1 and run 2 and theoretically calculated for zircon-type phase at ambient pressure and scheelite-type phase.

| | | Zircon | | | | |
|---|---|---|---|---|---|---|
| | | Atom | Site | X | y | z |
| | | Sm | 4a | 0 | 3/4 | 1/8 |
| | | V | 4b | 0 | 1/4 | 3/8 |
| Run 1 | | O | 16h | 0 | 0.435(2) | 0.207(2) |
| Run 2 | | O | 16h | 0 | 0.432(1) | 0.203(1) |
| Theory | | O | 16h | 0 | 0.43132 | 0.20308 |
| | | Scheelite | | | | |
| | | Sm | 4b | 0 | 1/4 | 5/8 |
| | | V | 4a | 0 | 1/4 | 1/8 |
| Run 1 (10 GPa) | | O | 16f | 0.252(3) | 0.063(2) | 0.069(2) |
| Run 2 (7.9 GPa) | | O | 16f | 0.251(1) | 0.110(1) | 0.0464(5) |
| Theory | 7.9 GPa | O | 16f | 0.25211 | 0.10792 | 0.04397 |
| | 12.3 GPa | | | 0.25181 | 0.10629 | 0.04337 |

Table II. EOS parameters for low and high pressure phases determined from experiments and calculations. For the calculations a 4$^{th}$ order EOS was used.

| Phase | PTM | $V_0$ (Å$^3$) | B (GPa) | B' | B'' (GPa$^{-1}$) |
|---|---|---|---|---|---|
| Zircon | MEW | 336.5(9) | 129(4) | 4 | - |
| Zircon | ME | 338.6(9) | 131(7) | 4 | - |
| Zircon | Theory | 346.5 | 118.2 | 5.1 | -0.0006 |
| Scheelite | MEW | 299.7(1.2) | 133(5) | 4 | - |
| Scheelite | ME | 298.9(1.5) | 256(12) | 4 | - |
| Scheelite | Theory | 306.8 | 139.7 | 3.9 | -0.06 |

Table III. Elastic constants at P = 0 GPa for the zircon-type and scheelite-type structure of $SmVO_4$.

| Elastic constant (GPa) | Zircon-type | Scheelite-type |
|---|---|---|
| $C_{11}$ | 212 | 218 |
| $C_{12}$ | 37 | 114.4 |
| $C_{13}$ | 76 | 99.5 |
| $C_{33}$ | 286.6 | 196.3 |
| $C_{44}$ | 40.8 | 53.4 |
| $C_{66}$ | 13.6 | 68.1 |
| $C_{16}$ | - | 21.8 |